\documentclass[aps,prd,preprint,showpacs,floats]{revtex4}
\usepackage[dvips]{graphicx}
\usepackage{bm}
\begin{document}
\title{$B \to \chi_{c0,2} K$ decays: a model estimation }
 \author{T. N. Pham}
 \email[E-mail address: ]{Tri-Nang.Pham@cpht.polytechnique.fr}
 \affiliation{Centre de Physique Theorique, \\
Centre National de la Recherche Scientifique, UMR 7644,\\
Ecole Polytechnique, 91128 Palaiseau Cedex, France}
  \author{Guohuai Zhu}
  \email[E-mail address: ]{Guohuai.Zhu@desy.de}
  \thanks{Alexander-von-Humboldt Fellow}
 \affiliation{ II. Institut f{\"u}r Theoretische Physik, Universit{\"a}t Hamburg, \\
Luruper Chaussee 149, 22761 Hamburg, Germany}
  \date{\today}
\begin{abstract}
    In this paper, we investigate the vertex corrections and
    spectator hard scattering contributions to $B \to \chi_{c0,2}K$
    decays, which has no leading contribution from naive factorization scheme.
    A non-zero binding energy $b=2m_c-M$ is introduced to regularize
    the infrared divergence of the vertex part. The spectator diagrams
    also contain logarithmic and linear infrared divergences, for which
    we adopt a model dependent parametrization. If we neglect possible strong
    phases in the hard spectator contributions , we obtain a too small
     branching ratio
    for $\chi_{c0}K$ while too large one for $\chi_{c2}K$, as can be seen
    from  the ratio of the branching ratio of $B^+ \to \chi_{c2}K^+$ to
    that of $B^+ \to \chi_{c0} K^+$, which is predicted to be
     $2.15^{+0.63}_{-0.76}$
    in our model, while experimentally it should be about $0.1$ or even
    smaller. But a closer examination shows that, assuming large strong
    phases difference between the twist-2 and twist-3 spectator terms, together
    with a slightly larger spectator infrared cutoff parameter $\Lambda_h$,
    it is possible to accommodate the experimental data. This shows
    that, for  $B\to \chi_{c0,2}K$ decays with no factorizable 
    contributions, QCDF
    seems capable of producing decay rates close to experiments, in 
    contrast
    to the $B\to J/\psi K$ decay which is dominated by the factorizable
     contributions.
\end{abstract}
\maketitle
\section{introduction}
Hadronic B decays attract a lot of attention because of its role in
determining the Cabibbo-Kobayashi-Maskawa (CKM) matrix elements,
extracting CP-violating angles and even revealing physics beyond the
Standard Model (SM). However in most cases, a deep understanding on
the strong dynamics in hadronic B decays is prerequisite for the
above purposes.

Phenomenologically the naive factorization ansatz (NF) \cite{BSW},
supported by color transparency argument \cite{Bjorken}, is widely
used in hadronic two-body B decays. However the unphysical
dependence of the decay amplitude on renormalization scale indicates
a prominent role of QCD corrections to NF. In this respect, $B \to
\chi_{c0,2}K$ decays are of special interest as these channels
vanish in the approximation of NF, due to the spin-parity and vector
current conservation. Therefore they provide a good opportunity to
study the QCD corrections to NF. It was generally believed that the
branching ratios of these channels should be quite small as the QCD
corrections are either suppressed by strong coupling $\alpha_s$ or
$\Lambda_{QCD}/m_b$. But BaBar \cite{Babar} and Belle \cite{Belle}
have found a surprisingly large branching ratio of $B^+ \to
\chi_{c0} K^+$ decay,
\begin{equation}
{\cal B}(B^+ \to \chi_{c0} K^+)=\left \{ \begin{array}{ll}
(6.0^{+2.1}_{-1.8}\pm 1.1) \times 10^{-4} & \mbox{(Belle)}~, \\
(2.7 \pm 0.7) \times 10^{-4} & \mbox{(BaBar)}~.
\end{array} \right.
\end{equation}
Actually this large branching ratio  is even comparable,
for example, to that of $B \to \chi_{c1} K$
decay which is not forbidden in NF. Another surprising observation
is that, the upper limit of $B \to \chi_{c2} K$ decay is roughly an
order of magnitude smaller than the observed branching ratio of $B^+
\to \chi_{c0} K^+$ decay \cite{Babar2},
\begin{equation}
{\cal B}(B^+ \to \chi_{c2} K^+)<3.0 \times 10^{-5} \hspace*{1cm}
\mbox{(BaBar)}~,
\end{equation}
while naively the branching ratios of $B \to \chi_{c0,2} K$ decays
are expected to be at the same order.

In the following we shall discuss these decay channels using the QCD
factorization (QCDF) approach \cite{BBNS}. In this framework, the
final state light meson is described by the light-cone distribution
amplitude(LCDA), while for the $P$-wave charmonium $\chi_{c0,2}$, we
shall adopt the covariant projection method of non-relativistic QCD
\cite{Petrelli}. It is well known that, for the inclusive decay and
production of $P$-wave charmonia, the color-octet mechanism must be
introduced to guarantee the infrared safety. However it is still
unclear how to incorporate this mechanism in a model-independent way
into exclusive processes. Thus the decay amplitudes ${\cal A}(B \to
\chi_{c0,2} K)$ would be inevitably infrared divergent when only the
color-singlet picture is adopted for $\chi_c$, which is shown
explicitly in \cite{Chao}. Thus strictly speaking, the QCDF approach
is not applicable for $B \to \chi_{c0,2} K$ decays due to the
breakdown of factorization.

In this paper, to get a model estimation, we will introduce the
binding energy $b=2m_c-M$ \cite{Kuhn} as an effective cut-off to
regularize the infrared divergence appearing in the diagrams of
vertex corrections (see Fig. 1). In fact, the logarithmic divergence
$\ln(b)$ term in the limit $b \to 0$  for the vertex corrections in
$B \to \chi_{c0,2} K$ decays is similar to the $\ln(b)$ term found
in Ref. \cite{Kuhn} for the production of $P$-wave charmonium in
$e^{+}e^{-}$ collisions. As for the spectator scattering
contributions, there appears logarithmic divergence at twist-2 level
and linear divergence at twist-3 level. Phenomenologically we shall
parameterize these divergence as $\ln[m_B/\Lambda_h]$ and
$m_B/\Lambda_h$ respectively, where the non perturbative parameter
$\Lambda_h=500~\,\rm MeV$ again acts as an effective cut-off to regularize
the endpoint divergence \cite{Zhu}. According to the QCDF approach, all other
contributions are power suppressed by $\Lambda_{QCD}/m_b$.

We find that, with the above method, the branching ratio of $B^+ \to
\chi_{c0} K^+$ decay is about $0.78 \times 10^{-4}$, which is
several times smaller than the experimental measurements. At the
same time, we also get the branching ratio of $B^+ \to \chi_{c2}
K^+$ decay at about $1.68 \times 10^{-4}$, which is significantly
larger than the upper limit $3 \times 10^{-5}$ observed by Babar
\cite{Babar2}. But the above estimation is very crude in that the
strong phases effects are completely ignored. Notice further that
for the spectator contributions, there contains only logarithmic
divergence at twist-2 level, while linear divergence appears at the
twist-3 level, the strong phases of the twist-2 and twist-3
spectator terms could be quite different. We then briefly discuss
the potential strong phases effects and argue that very different
strong phases between twist-2 and twist-3 spectator terms together
with a slightly larger $\Lambda_h$ seems to be able to reproduce the
experimental hierarchy ${\cal B}(B^+ \to \chi_{c0} K^+) \gg {\cal
B}(B^+ \to \chi_{c2} K^+) $.

\begin{figure}[htb]
\begin{center}
\unitlength 1mm
\begin{picture}(150,90)
\put(0,0){\includegraphics{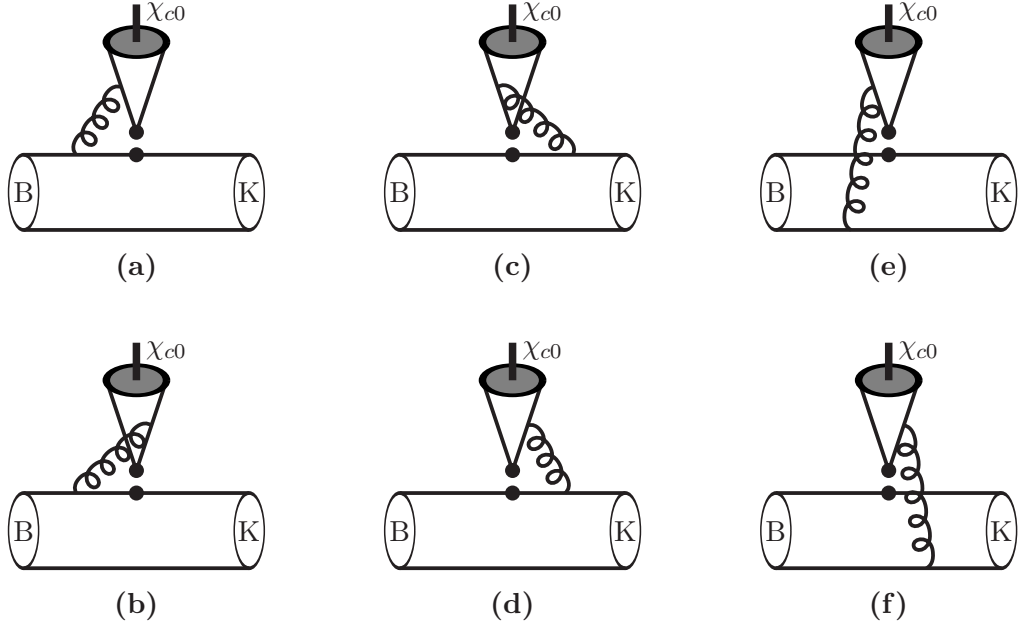}}
\end{picture}
\vspace*{-0.5cm} \caption{Order of $\alpha_s$ contributions to
 $B \to \chi_{cJ}K$ decay. Fig. (a)-(d) and Fig. (e)-(f) are called vertex
 corrections and spectator scattering diagrams, respectively.}
\end{center}
\end{figure}

\section{Vertex and spectator corrections}
In the QCDF approach, $K$ meson is described by the following
light-cone projection operator in momentum space\cite{BBNS}
\begin{equation}\label{projector}
M^K_{\alpha \beta} = \frac{if_K}{4} \left \{ \slash{\hskip -2mm}l
\gamma_5 \Phi(x) - \mu_K \gamma_5 \frac{\slash{\hskip -1.7mm}l_2
\slash{\hskip -1.7mm} l_1}{l_2 \cdot l_1}\Phi_P(x) \right \}_{\alpha
\beta }~,
\end{equation}
where $l$ is the momentum of $K$ meson and $l_1$($l_2$) is the
momentum of quark (antiquark) in K meson. $\Phi(x)$ and $\Phi_P(x)$
are leading twist and twist-3 distribution amplitudes of $K$ meson,
respectively. It is understood that only after the factor $l_2 \cdot
l_1$ in the denominator is canceled, may we take the collinear limit
$l_1=xl$, $l_2=(1-x)l$. Notice that in principle we could also start
directly from the original light-cone projector of K meson in
coordinate space \cite{Braun}, and the physical results should be
the same. But in this case care must be taken that, only with a
proper regularization, can one do the relevant convolution integrals
correctly. The readers may refer to the appendix of \cite{twist3}
for further details.

  Since $P$-wave charmonium $\chi_{cJ}$ is involved, we shall
use covariant projection method \cite{Petrelli,Kuhn} to calculate
the decay amplitude
\begin{equation}
{\cal A}(B\to\chi_{c0,2}K)={\cal E}^{(0,2)}_{\alpha \beta}
\frac{\partial}{\partial q_\beta}Tr \left [ \Pi^\alpha_1 {\cal C}_1
{\cal A} \right ]|_{q=0}~.
\end{equation}
Here ${\cal A}$ is the standard QCD amplitude for $c {\bar c}$
production, amputated of the heavy quark spinors, ${\cal
C}_1=\delta_{ij}/\sqrt{3}$ is the color singlet projector. While
$\Pi^\alpha_1$ is the $S=1$ heavy quark spinor projector
\begin{equation}
\Pi^\alpha_1 = \frac{1}{\sqrt{8m_c^3}}\left ( \frac{\slash{\hskip
-2.5mm}P}{2}-\slash{\hskip -2mm}q-m \right ) \gamma^\alpha \left (
\frac{\slash{\hskip -2.5mm}P}{2}+\slash{\hskip -2mm}q+m \right )~,
\end{equation}
where $P$ is the momentum of charmonium and $2q$ is the relative
momentum between the $c{\bar c}$ pair in $\chi_{cJ}$. ${\cal
E}^{(0,2)}_{\alpha \beta}$ is the polarization tensor of
$\chi_{c0,2}$ which satisfies the following sum over polarization
relation \cite{Petrelli}
\begin{eqnarray}
{\cal E}^{(0)}_{\alpha \beta} {\cal E}^{(0)}_{\alpha' \beta'} &=&
\frac{1}{D-1} \Pi_{\alpha \beta} \Pi_{\alpha' \beta'}~, \nonumber \\
{\cal E}^{(2)}_{\alpha \beta} {\cal E}^{(2)}_{\alpha' \beta'} &=&
\frac{1}{2} \left ( \Pi_{\alpha \alpha'} \Pi_{\beta \beta'}+
                    \Pi_{\alpha \beta'}  \Pi_{\beta \alpha'}
                    \right ) - \frac{1}{D-1} \Pi_{\alpha \beta}
                    \Pi_{\alpha' \beta'}~,
\end{eqnarray}
with
\begin{equation}
\Pi_{\alpha \beta}=-g_{\alpha \beta} + \frac{P_\alpha
P_\beta}{M^2}~.
\end{equation}
Here M is the mass of $\chi_{cJ}$.

For charmonium B decays, we shall start with the effective
Hamiltonian \cite{Buras}
\begin{equation}
{\cal H}_{eff}=\frac{G_F}{\sqrt{2}} \left \{ V_{cb}V_{cs}^*
(C_1(\mu)Q_1^c(\mu)+C_2(\mu)Q_2^c(\mu))-V_{tb}V_{ts}^*
\sum_{i=3}^{6} C_i(\mu)Q_i(\mu) \right \}~,
\end{equation}
where $C_i$ are Wilson coefficients which are perturbatively
calculable and $Q_{1,2}$($Q_{3-6}$) are the effective tree (QCD
penguin) operators. Notice that we have dropped the electroweak
penguin contributions here which are numerically negligible. The
four-quark effective operators are defined as
\begin{equation}
\begin{array}{l}
\begin{array}{ll}
Q^c_1= ( \bar{q}_{\alpha} b_{\alpha} )_{V-A}
         ( \bar{c}_{\beta} c_{\beta} )_{V-A}~,&
Q^c_2= ( \bar{s}_{\alpha} b_{\beta} )_{V-A}
         ( \bar{c}_{\beta} c_{\alpha} )_{V-A}~,\\
Q_{3,5}= (\bar{s}_{\alpha} b_{\alpha} )_{V-A}
      \sum\limits_{q}
     ( \bar{q}_{\beta} q_{\beta} )_{V \mp A}~,~~~~~~&
Q_{4,6}= (\bar{s}_{\beta} b_{\alpha} )_{V-A}
      \sum\limits_{q}
     ( \bar{q}_{\alpha} q_{\beta} )_{V \mp A}~.\\
\end{array} \\
\end{array}
\end{equation}
Here $q$ denotes all the active quarks at the scale $\mu={\cal
O}(m_b)$, i.e. q=u, d, s, c, b. While $\alpha$ and $\beta$ are color
indices.

It is then straightforward to get the decay amplitude of $B \to
\chi_{c0,2} K$ decay by considering the vertex and spectator
corrections drawn in Fig. 1,
\begin{eqnarray}\label{am}
{\cal A}(B \to \chi_{c0,2} K)&=&\frac{i G_F}{\sqrt{2}}
\frac{6|R'_1(0)|}{\sqrt{\pi M}}\frac{\alpha_s}{4\pi}\frac{C_F}{N_c}
(V_{cb}V_{cs}^* C_1-V_{tb}V_{ts}^* (C_4+C_6)) \times \nonumber \\
&& F_0^{B\to K} \left ( f^I_{(0,2)}+\frac{4\pi^2}{N_c}\frac{f_B
f_K}{F_0^{B\to K}} (f_{(0,2)}^{II2}+f_{(0,2)}^{II3}) \right )~,
\end{eqnarray}
where $R'_1(0)$ is the derivative of the $\chi_{cJ}$ wave function
at the origin and $F_0^{B\to K}$ the form factor of $B\to K$. The
function $f^I$ represents the contributions from vertex corrections
while $f^{II2}$($f^{II3}$) arising from the twist-2 (twist-3)
spectator contributions. The vertex function $f^I$ is actually
infrared divergent and therefore depends on the binding energy
$b=2m_c-M$. In the following we shall keep $\ln(b/M)$ term and drop
the terms suppressed by $b/M$. The explicit expressions of
$f^I_{(0,2)}$ are as follows
\begin{eqnarray}\label{fI}
f^I_0&=&\frac{2 m_B ((1+12a)(1-4a)+16a
\ln{[4a]})}{(1-4a)^2\sqrt{3a}} \ln{\left [\frac{-b}{M} \right
]}+f^I_{0\,\rm fin}+{\cal O}(b/M)~, \nonumber \\
f^I_2&=&\frac{32 {\cal E}^{(2)*}_{\alpha \beta} p_B^\alpha p_B^\beta
\sqrt{a} ((1+12a)(1-4a)+16a \ln{[4a]})}{m_B(1-4a)^3} \ln{\left
[\frac{-b}{M} \right ]}+f^I_{2\,\rm fin}+{\cal O}(b/M)~,
\end{eqnarray}
where $a=m_c^2/m_b^2$ , and  $f^I_{\rm fin}$ is the finite part of the
function $f^I$ in the limit $b/M\to 0$.  The explicit expressions of
$f^I_{\rm fin}$ are as follows,
\begin{eqnarray}
f^I_{0\,\rm fin}&=&\frac{-m_B}{2(1-4a)^2(1-2a)^3 \sqrt{3a}}\left \{ -6 -
22\ln{2} +
  4a(26 + (15 - 56\ln{2})\ln{2} \frac{}{}+\right . \nonumber \\
    && 8a^2 (65 + 52\ln{2} - 84\ln^2{2})  +
    384 a^4 (1 + 2\ln{2}) +
    2a (-85 + 28\ln{2}(-1 + 6\ln{2}))+ \nonumber \\
    && 32a^3 (-23 - 32\ln{2} + 14\ln^2{2})) -
  8\ln{a} + 4a(-(1 - 4a)^2 (5 - 24(1 - a)a) \times \nonumber \\
    &&  \ln{\left [\frac{-1 + 4a}{a}\right ]} +
      9\ln{a} + 2
     (a(-3 + 4a)(13 - 46a + 56 a^2) - \nonumber \\
    &&
       4(1 - 2a)^3
       \ln{[64a]})\ln{a} + 16(1 - 2a)^3\ln{2}
     \ln{[-1 + 4a]}) - \nonumber \\
    && \left .
       64a(1 - 2a)^3
   \left ( \mbox{Li2}\left [ \frac{2-4a}{1 - 4a}\right ] +
    \mbox{Li2}[1 - 4a] - \mbox{Li2}\left [
     \frac{1 - 2a}{1 - 4a} \right ] \right )
\right \}~, \nonumber \\
f^I_{2\,\rm fin}&=&\frac{-{\cal E}^{(2)*}_{\alpha \beta} p_B^\alpha
p_B^\beta}{4m_B \sqrt{a}} \left \{  \frac{32a}{(1-2a)^3(1-4a)}
\left (\frac{}{} 4\ln{2}(1-2a)^2+(1 - 4 a) (4 a^2 (1 + 2\ln{2}) - 1)
\right. \right. \nonumber \\
&&\hspace*{2cm} \left. \left.  - 8 a (3 a - 1) (\ln{a} - \ln{[4 a -
1]})\frac{}{}\right )+V_{AB}[a] \right \}~,
\end{eqnarray}
where the function $V_{AB}[a]$ denotes the finite part of vertex
corrections from Fig. (1a-1b).  The analytical form of $V_{AB}[a]$
is too complicated to be shown here, but numerically it has a very
mild dependence on the parameter a, for example,
\[  V_{AB}[0.1]=11.3~, \hspace*{1.5cm} V_{AB}[0.15]=11.9~.\]

As for the spectator functions, we have
\begin{eqnarray}\label{fII}
f_0^{II2}&=&\frac{1}{m_b(1-4a)\sqrt{3a}} \int_0^1 d\xi
\frac{\phi_B(\xi)}{\xi} \int_0^1 dy
\frac{\phi_K(y)}{\bar{y}^2}(-8a+(1-4a)\bar{y})~, \nonumber \\
f_0^{II3}&=&\frac{2}{m_b(1-4a)^2\sqrt{3a}} \frac{\mu_K}{m_b}
\int_0^1 d\xi \frac{\phi_B(\xi)}{\xi} \int_0^1 dy
\frac{\phi_P(y)}{\bar{y}^2}(8a-(1-4a)\bar{y})~, \nonumber \\
f_2^{II2}&=&\frac{16{\cal E}^{(2)*}_{\alpha \beta} p_B^\alpha p_B^\beta
\sqrt{a}}{m_b^3 (1-4a)^3} \int_0^1 d\xi \frac{\phi_B(\xi)}{\xi}
\int_0^1 dy
\frac{\phi_K(y)}{\bar{y}^2}(4a+(1-4a)\bar{y})~, \nonumber \\
f_2^{II3}&=&\frac{32{\cal E}^{(2)*}_{\alpha \beta} p_B^\alpha p_B^\beta
\sqrt{a}}{m_b^3 (1-4a)^4}\frac{\mu_K}{m_b} \int_0^1 d\xi
\frac{\phi_B(\xi)}{\xi} \int_0^1 dy
\frac{\phi_P(y)}{\bar{y}^2}(8a-(1+8a)\bar{y})~.
\end{eqnarray}
Here $\xi$ is the momentum fraction of the light spectator quark in
the $B$ meson, and $\bar{y}=1-y$ the light-cone momentum fraction of
the quark in the $K$ meson which is from the spectator quark of $B$
meson. Notice that our expressions for twist-2 spectator function
$f_{(0,2)}^{II2}$ are consistent with those of \cite{Chao}.

\section{Numerical results and discussion}
To get a numerical estimation on the branching ratios of $B \to
\chi_{c0,2} K$ decays, several parameters appearing in Eqs.
(\ref{am} $\sim$ \ref{fII}) should be first decided on. The
derivative of $\chi_{cJ}$ wave function at the origin $|R'(0)|$ may
be either estimated by QCD-motivated potential models \cite{Quigg},
or extracted from $\chi_{cJ}$ decays \cite{Mangano}. $|R'(0)|^2$
varies from $0.075~\,\rm GeV^5$ to $0.131~\,\rm GeV^5$ in different potential
models \cite{Quigg} while using $\chi_{cJ}$ decays, for instance
\cite{Barbieri},
\begin{equation}
\Gamma(\chi_{c2} \to \gamma \gamma)=\frac{36}{5}e_Q^4 \alpha_{em}^2
\frac{|R'(0)|^2}{m_c^4} \left ( 1-\frac{16}{3}\frac{\alpha_s}{\pi}
\right )~,
\end{equation}
it is easy to get $|R'(0)|^2=(0.062 \pm 0.007)~\, \rm GeV^5$ if we take
$m_c=1.5~\, \rm GeV$. This result is a little bit lower than, but still
consistent with the potential model calculations, especially
considering that it is very sensitive to the choice of charm quark
mass.  In this paper, we shall take $|R'(0)|^2=(0.10 \pm 0.03)~
\,\rm GeV^5$ as input.

For the binding energy, if we take $m_c=1.5~\,\rm GeV$, the ratio $b/M$ is
about $-0.11(-0.16)$ for $\chi_{c0}$($\chi_{c2}$), while
$a=m_c^2/m_b^2 \simeq 0.1$. The QCD scale $\mu$ should be order of
$\sqrt{m_b \Lambda}$, as in charmless $B$
decays, which is about $(1 \sim 1.5)~\,\rm GeV$. In the following we shall
fix the scale $\mu=1.3~\,\rm GeV$ with $\alpha_s=0.36$. Notice also that
the Wilson coefficients should be evaluated at leading order, to be
consistent with the leading order formula of Eq. (\ref{am}),
\begin{equation}
C_1=1.26~, \hspace{0.6cm} C_4=-0.049~, \hspace{0.6cm} C_6=-0.074~.
\end{equation}
The relevant CKM parameters are chosen to be $A=0.83$ and
$\lambda=0.224$.

As for the spectator contributions, we adopt the following LCDAs for
the final $K$ meson,
\begin{equation}
\phi_K(y)=6y(1-y)(1+\sum_{n \ge 1} a_n C_n^{(3/2)} (2y-1)~,
\hspace{1cm} \phi_P(y)=1~,
\end{equation}
where $C^{(3/2)}_n(x)$ are Gegenbauer polynomials. The parameters
$a_n$ are set to be \cite{Ball}:
\begin{equation}
 a_1=0.17~,\hspace{0.5cm} a_2=0.115~,\hspace{0.5cm}
 a_4=0.015~,\hspace{0.5cm} a_3=a_{n>4}=0~.
\end{equation}

Then logarithmic and linear divergences appear in Eq.(\ref{fII}),
which may be phenomenologically parameterized as \cite{BBNS}
\begin{equation}\label{endpoint}
\int \frac{dy}{y}=\ln \frac{m_B}{\Lambda_h}~, \hspace{1cm} \int
\frac{dy}{y^2}=\frac{m_B}{\Lambda_h}~,
\end{equation}
with $\Lambda_h=500~\,\rm MeV$. Notice that the above
parametrization of linear divergence would violate the power
counting of QCDF, but we do not have better way yet to deal with it.
This is clearly a very rough estimation, for example, we do not
consider here the strong phase effect. We also know little about $B$
wave function, but fortunately only the following integral is
involved which may be parameterized as
\begin{equation}
\int d\xi \frac{\phi_B(\xi)}{\xi}=\frac{m_B}{\lambda_B}~.
\end{equation}
and we shall simply fix $\lambda_B=350~\,\rm MeV$ in our
calculation. The chirally enhanced ratio $r_K=\mu_K/m_b$ is chosen
to be $0.43^{+0.11}_{-0.08}$, which corresponds to taking $m_s(2~\rm
GeV )=(90 \pm 20)~\rm MeV $ and $(m_u+m_d)(2~\rm GeV )=9~\rm MeV$.
The form factor $F_0^{B\to K}(m_{\chi_c}^2)$ may be read from
\cite{Ball}, in which as stated, the uncertainty of form factor at
$q^2 \ne 0$ is likely to be smaller than that of $q^2=0$, which is
about $12\%$. Therefore we will cite $F_0^{B\to
K}(m_{\chi_c}^2)=0.48 \pm 0.06$ as our input. The decay constants
are set as $f_K=160~\,\rm MeV$ and $f_B=(210 \pm 25)~\,\rm MeV$.
With the above input, we get
\begin{eqnarray}
{\cal B}(B^+ \to \chi_{c0} K^+)&=&(0.78^{+0.46}_{-0.35}) \times
10^{-4}~,\nonumber \\
{\cal B}(B^+ \to \chi_{c2} K^+)&=&(1.68^{+0.78}_{-0.69}) \times
10^{-4}~.
\end{eqnarray}
We also show separately the contributions from vertex corrections
and hard spectator scattering diagrams in Table I, with all the
input parameters taken at their central values.
\begin{table}[htb]
\begin{center}
\caption{The numerical estimations of vertex corrections and hard
spectator scattering contributions, with all the parameters taken at
their central values. The constant $C \equiv \frac{4\pi^2}{N_c}
\frac{f_Bf_K}{F_0^{B->K}}$.}
\begin{tabular}{|c|c|c|c|}\hline
Decay Channels & $f^I$ & $C*f^{II2} $ & $C*f^{II3} $ \\ \hline

$\chi_{c0}K$ & $46.3-33.6 i$  & -43.1  & 80.7 \\ \hline

$\chi_{c2}K$ & $1.7+14.1 i$   & 69.3   & 68.3 \\ \hline
\end{tabular}
\end{center}
\end{table}
For the case of $\chi_{c0}K$ channel, our results are approximately
four times smaller than the average of BaBar and Belle measurements,
$(3.0 \pm 0.7) \times 10^{-4}$, while our prediction on $B \to
\chi_{c2}K$ decay is obviously too large compared with the
experimental upper limit, $3.0 \times 10^{-5}$. This is a little bit
surprising, because for charmonium B decays, the theoretical results
are normally a few times {\it smaller} than the experimental
measurements.

The careful reader may have noticed that in the above analysis we
did not consider the uncertainty related to the parameter
$a=m_c^2/m_b^2$. In fact a larger $a$ could enhance the branching
ratio of $B \to \chi_{c0}K$ decay significantly, but unfortunately
it would also enhance that of $B \to \chi_{c2} K$ decay with similar
magnitude. Notice that $B \to \chi_{c0,2} K$ share many common
inputs, the ratio of branching ratios of these two channels should
have mild dependence on the input parameters, for example it is
independent on the parameter $|R'(0)|$. Our numerical analysis shows
that this is indeed the case, with $a=0.10 \pm 0.03$:
\begin{equation}\label{ratio}
{\cal R}=\frac{{\cal B}(B^+ \to \chi_{c2} K^+)}{{\cal B}(B^+ \to
\chi_{c0}
K^+)}=2.15^{+0.26}_{-0.55}~^{+0.33}_{-0.31}~^{+0.36}_{-0.28}~^{+0.30}_{-0.31}~,
\end{equation}
where the uncertainties arise from the parameters $a$, $r_K$,
$F^{B->K}_0$ and $f_B$, respectively. The above ratio is clearly in
strong contradiction with the experimental hierarchy ${\cal R}
\lesssim 0.1 \ll 1$.

Notice that the chirally enhanced power corrections, namely twist-3
spectator contributions in this case, have been included in the
above estimation. For the rest part of power corrections, there is
no systematic way to estimate them yet. But since the power corrections
are suppressed by $\Lambda_{QCD}/m_b$, intuitively they might lead
to an uncertainty of about $20\%$ to the decay amplitude, which is
unlikely to be able to change our estimation Eq.(\ref{ratio})
dramatically.

In our model, the parameters $\Lambda_h$ and $\lambda_B$ will
introduce additional uncertainties to $B \to \chi_{c0,2} K$ decays.
It is very unlikely that we could reproduce the experimental
observations by fine tuning $\lambda_B$, because although a larger
$\lambda_B$ would lead to a smaller branching ratio for $\chi_{c2}
K$ decay, it would also make the already too small branching ratio
of $\chi_{c0} K$ decay even smaller. However a larger $\Lambda_h$
does help to close the gap between our predictions and the
experimental data, due to the fact that a larger $\Lambda_h$ will
lead to a significantly smaller branching ratio for $\chi_{c2} K$
decay while $\chi_{c0} K$ decay does not change much. Of course we
can not choose a too large $\Lambda_h$, say larger than $1~\,\rm
GeV$, because it is anyway a non perturbative parameter. As an
illustration, we take $\Lambda_h=700~\,\rm MeV$ and get
\begin{equation}
{\cal B}(B^+ \to \chi_{c0} K^+)=0.78 \times 10^{-4}~,\hspace{0.6cm}
{\cal B}(B^+ \to \chi_{c2} K^+)=0.74 \times 10^{-4}~.
\end{equation}
Although it seems to be on the right way, this effort alone is still
not enough to accommodate the experimental data.

Let us take a closer look at the decay amplitudes. From Table I, it
is clear that the spectator hard scattering mechanism is dominant in
$B \to \chi_{c2} K$ decay and also very important for $\chi_{c0}K$
channel. Furthermore there is significant destructive (constructive)
interference between the twist-2 spectator term and the twist-3 one
for $\chi_{c0} K$ ($\chi_{c2} K$) mode. This is probably the reason
that we get too small $\chi_{c0}K$ decay as well as too large
$\chi_{c2}K$ decay in our model. Notice that there are logarithmic
and linear divergences appear in the spectator contributions, which
are parameterized by Eq. (\ref{endpoint}). It is obviously a very
rough model estimation and for example, strong phases effects are
completely ignored. It is also reasonable to assume that the strong
phase of twist-2 spectator term could be different from that of
twist-3 part. As an illustration, the endpoint divergences could be
parameterized as \cite{BBNS}:
\begin{equation}
\int \frac{dy}{y}=\ln \frac{m_B}{\Lambda_h}(1+\rho_{2,3} e^{i
\theta_{2,3}})~, \hspace{1cm} \int
\frac{dy}{y^2}=\frac{m_B}{\Lambda_h}(1+\rho_3 e^{i \theta_3})~,
\end{equation}
with $0 \leq \rho \leq 1$ and the phase $\theta$ completely free. In
the above equations, $(\rho_2,\theta_2)$ denotes the parameters for
twist-2 spectator term and $(\rho_3,\theta_3)$ for twist-3 one. In
this case, the interference effects and therefore the predictions of
the branching ratios, could be changed dramatically. For example, if
we take a somewhat extreme case
\[
 \rho_2=0.6~,\hspace*{1cm} \theta_2=\pi~,\hspace*{1cm}
 \rho_3=0~,\hspace*{1cm} \theta_3=0~,
 \]
with $\Lambda_h=600~\,\rm MeV$ while keep all other input parameters
fixed at their central values, we will get
\begin{equation}
{\cal B}(B^+ \to \chi_{c0} K^+)=3.3 \times 10^{-4}~,\hspace{0.6cm}
{\cal B}(B^+ \to \chi_{c2} K^+)=1.7 \times 10^{-5}~,
\end{equation}
which are in good agreement with the experimental observations.
Certainly, due to the non perturbative nature of the above strong
phases, there is strong model dependence of our predictions.
Therefore it is not so meaningful to fine tune the parameters to get
the best fit of the experimental data. The key point here is that,
different strong phases between twist-2 and twist-3 spectator terms
might be able to account for the experimental hierarchy that ${\cal
B}(B \to \chi_{c0}K)$ is at least an order of magnitude larger than
${\cal B}(B \to \chi_{c2}K)$.

The authors of ref. \cite{chao2} also studied the $B \to \chi_{c0}
K$ decay with the same QCDF method. But they used the gluon mass and
gluon momentum cutoff, instead of binding energy adopted in this
paper, to regularize the infrared divergences of the vertex
corrections. Another difference is that they calculated the
spectator contributions directly from the original light-cone
projector of K meson in coordinate space and got a different result
from this paper. They claimed that the difference was due to the
light-cone projector adopted in this paper which is inappropriate
for $\chi_{cJ} K$ channels: to get the projector
Eq.(\ref{projector}) from the original one in coordinate space
\cite{Braun}, the integration by parts has been used and the
boundary terms were dropped. However because of the linear
singularities appeared in the above calculations, the boundary terms
seem to be divergent and thus the justification of using the
integration by parts is in doubt in this case. But Beneke has
elaborated on this subtle point in the appendix of ref.
\cite{twist3} and it is shown there that the boundary terms are
indeed zero provided the propagators are regularized carefully when
they go close to the mass-shell. Therefore the integration by parts
can be used here and the light-cone projector adopted in this paper
is justified. Certainly, with a proper regularization, the
calculation starting directly from the coordinate space projector
should give the same results as this paper.

Most recently, the $B \to \chi_{c0} K$ decay was discussed by using
the PQCD method \cite{Li}. Notice that the vertex corrections were
not included in their calculations, and the spectator contributions
alone are enough in their paper to account for the experimental
data. It would be very interesting to see whether they could also
reproduce the very small branching ratio of $\chi_{c2} K$ channel
observed by Babar, which has not been done yet.

The $B \to \chi_{c0} K$ decay was also analyzed with light-cone sum
rules \cite{Huang,Melic}. Although there are some discrepancies in
their papers, they agreed on the point that their results were too
small to accommodate the experimental data. A large charmed meson
rescattering effects $B \to D_s^{(*)}D^{(*)} \to \chi_{c0,2} K$
could account for the surprisingly large $B \to \chi_{c0} K$
decay\cite{Pham}, but generally it will also lead to a large
branching ratio for $\chi_{c2} K$ mode.

In summary, we discuss in this paper the vertex corrections and
spectator hard scattering contributions to $B \to \chi_{c0,2} K$
decays. Since there is no model independent way yet to estimate the
color-octet contribution to exclusive processes, it is no wonder
that the vertex corrections here are infrared divergent. The
non-zero binding energy $b=2m_c-M_{\chi_{cJ}}$ makes the charm quark
slightly off-shell inside $\chi_{cJ}$, and effectively acts as a
cut-off to regularize the vertex part. There are also less serious
logarithmic and linear endpoint divergences which appears in the
spectator contributions and are parameterized in a model-dependent
way as usually done in charmless B decays. This means that the
spectator diagrams are actually dominated by soft gluon exchange,
which in a sense could be viewed as a model estimation of
color-octet contributions. Then our numerical analysis predicts the
branching ratio of $B^+ \to \chi_{c0} K^+$ decay to be about $0.78
\times 10^{-4}$, about four times smaller than the experimental
observations, while for $B^+ \to \chi_{c2} K^+$ decay, we get $1.68
\times 10^{-4}$, which is about five times larger than the
experimental upper limit. But concerning the large theoretical
uncertainties, it is more interesting to consider the ratio $R={\cal
B}(B^+ \to \chi_{c2} K^+)/{\cal B}(B^+ \to \chi_{c0} K^+)$, in which
a large part of the theoretical uncertainty can be eliminated.
Numerically we find the ratio to be $R=2.15^{+0.63}_{-0.76}$, in
sharp contrast to the experimental observation that this ratio
should be about $0.1$ or even smaller, if the BaBar analysis of the
upper limit of ${\cal B}(B^+ \to \chi_{c2}K^+)$ decay will be
confirmed by farther measurements. We then have a closer look at the
decay amplitudes. One observation is that, $\chi_{c0} K$ channel is
not very sensitive to the spectator infrared cutoff parameter
$\Lambda_h$ while a larger $\Lambda_h$ could reduce the branching
ratio of $\chi_{c2} K$ decay significantly. Another observation is
that, in our model there is large destructive (constructive)
interference between the twist-2 and twist-3 spectator terms for
$\chi_{c0} K$ ($\chi_{c2} K$) mode. But notice that the twist-2
spectator contributions contain only logarithmic endpoint divergence
while twist-3 ones contain more severe linear endpoint divergence,
it should be reasonable to assume that their strong phases could be
quite different. Since the interference effects are very sensitive
to the strong phases difference, this might change our model
predictions dramatically. As an illustration, we then show in an
explicit case that, with a slightly larger $\Lambda_h$ and large
strong phases difference between twist-2 and twist-3 spectator
terms, our predictions are in good agreement with the experimental
data.

 In conclusion, what we have shown in this paper, is that by adjusting
the parameters for the spectator hard scattering  contributions, as
with the annihilation terms for charmless $B$ decays, QCDF is able
to produce appreciable non factorizable contributions to $B^+ \to
\chi_{c0,2}K^+ $ decays close to experiments, in contrast with the
$B^+ \to J/\psi K^+ $ decay which needs a large factorizable
contribution in addition to the small non factorizable one obtained
in QCDF \cite{Cheng}.


\section*{Acknowledgement}
G.Z acknowledges the support from Alexander-von-Humboldt Stiftung.


\begin{thebibliography}{99}
\bibitem{BSW}
M. Bauer, B. Stech and M. Wirbel, Z. Phys. {\bf C 34}, 103 (1987).

\bibitem{Bjorken}
J.D. Bjorken, Nucl. Phys. Proc. Suppl. {\bf 11}, 325 (1989).

\bibitem{Babar}
BaBar Collaboration, B. Aubert {\it et al.}, Phys. Rev. D {\bf 69},
071103 (2004).

\bibitem{Belle}
Belle Collaboration, K. Abe {\it et al.}, Phys. Rev. Lett. {\bf 88},
031802 (2002).

\bibitem{Babar2}
BaBar Collaboration, B. Aubert {\it et al.}, Phys. Rev. Lett. {\bf 94},
171801 (2005),hep-ex/0501061.

\bibitem{BBNS}
M. Beneke, G. Buchalla, M. Neubert and C.T. Sachrajda, Phys. Rev.
Lett. {\bf 83}, 1914 (1999); Nucl. Phys. B {\bf 591}, 313 (2000);
{\it ibid.}, {\bf 606}, 245 (2001).

\bibitem{Petrelli}
A. Petrelli {\it et al.}, Nucl. Phys. B {\bf 514}, 245 (1998).

\bibitem{Chao}
Z. Song, C. Meng, J.Y. Gao and K.T. Chao, Phys. Rev. D {\bf 69},
054009 (2004); Z. Song and K.T. Chao, Phys. Lett. B {\bf 568}, 127
(2003).

\bibitem{Kuhn}
J. Kuhn, J. Kaplan and E. Safiani, Nucl. Phys. B {\bf 157}, 125
(1979).

\bibitem{Zhu}
D.S. Du, D.S. Yang and G.H. Zhu, Phys. Lett. B {\bf 509}, 263 (2001).

\bibitem{Braun}
V.M. Braun and I.E. Filyanov, Z. Phys. C {\bf 44}, 157 (1989); Z.
Phys. C {\bf 48}, 239 (1990).

\bibitem{twist3}
M. Beneke, Nucl. Phys. Proc. Suppl. {\bf 111}, 62 (2002).

\bibitem{Buras}
G. Buchalla, A.J. Buras and M.E. Lautenbacher, Rev. Mod. Phys. {\bf
68}, 1125 (1996).

\bibitem{Quigg}
E. Eichten and C. Quigg, Phys. Rev. D {\bf 52}, 1726 (1995).

\bibitem{Mangano}
M. Mangano and A. Petrelli, Phys. Lett. B {\bf 352}, 445 (1995).

\bibitem{Barbieri}
R. Barbieri {\it et al.}, Nucl. Phys. B {\bf 192}, 61 (1981).

\bibitem{Ball}
P. Ball and R. Zwicky, Phys. Rev. D {\bf 71}, 014015 (2005).

\bibitem{chao2}
C. Meng, Y.J. Gao and K.T. Chao, hep-ph/0502240.

\bibitem{Huang}
Z.G. Wang, L. Li and T. Huang, Phys. Rev. D {\bf 70}, 074006 (2004).

\bibitem{Melic}
B. Melic, Phys. Lett. B {\bf 591}, 91 (2004).

\bibitem{Pham}
P. Colangelo, F. De Fazio and T.N. Pham, Phys. Lett. B {\bf 542}, 71
(2002).

\bibitem{Li}
C.H. Chen and H-n. Li, hep-ph/0504020.

\bibitem{Cheng}
H. Y. Cheng and K. C. Yang, Phys. Rev. D {\bf 63}, 074011 (2001).

\end{thebibliography}
\end{document}